\documentclass[aps,prb,twocolumn]{revtex4} 

\usepackage{graphicx}
\usepackage{dcolumn}
\usepackage{bm}
\usepackage{amsmath} 

\newcommand{\comment}[1]{}

\begin{document}

\title{The Two-Fluid Theory for Superfluid Hydrodynamics
and Rotational Motion}


\author{Phil Attard}
\affiliation{ {\tt phil.attard1@gmail.com}  7, 22 May, 10, 16, 26 Jun 2025}



\begin{abstract}
The two-fluid theory for superfluid hydrodynamics
is derived from  the fountain pressure result
that condensed bosons move at constant entropy,
meaning that superfluid flow is driven by the chemical potential gradient.
Explicit results for $^4$He below the $\lambda$-transition
show that the superfluid rotates,
which is consistent with measured data
but which contradicts Landau's irrotational principle.
\end{abstract}


\maketitle

%
\section{Introduction}
\setcounter{equation}{0} \setcounter{subsubsection}{0}
\renewcommand{\theequation}{\arabic{section}.\arabic{equation}}
%

\subsection{Two-Fluid Theory}

The primary purpose of this paper is to derive from first principles
the two-fluid equations for superfluid hydrodynamics.
The second point is to show explicitly that superfluid flow
is rotational flow,
in contradiction to the widely accepted principle
that superfluid flow is irrotational (Landau 1941).

The two-fluid equations, due to Tisza (1938), are
\begin{equation} \label{Eq:TwoFluid0}
m n_0 \frac {\partial {\bf v}_0}{\partial t}
=
\frac {-n_0}{n} \nabla p + \frac{n_0}{n} \sigma \nabla T ,
\end{equation}
and
\begin{equation} \label{Eq:TwoFluid*}
m n_* \frac {\partial {\bf v}_*}{\partial t}
=
\frac {-n_*}{n} \nabla p - \frac{n_0}{n}  \sigma \nabla T
+ \eta \nabla^2 {\bf v}_* .
\end{equation}
(The reader is warned that these equations appear in the literature
with myriad typographic errors.)

Here the subscript 0 denotes the condensed bosons,
often called the superfluid and denoted with subscript s;
they are also called He~II.
The subscript * denotes uncondensed bosons,
often called the normal fluid and denoted with subscript n;
they are also called He~I.
(Actually, sometimes helium~II  refers to the mixture
of normal fluid and superfluid below the $\lambda$-transition.)
Originally, the $\lambda$-transition in liquid helium-4
was identified with Bose-Einstein condensation by F. London (1938),
who, following Einstein (1924, 1925),
modeled the condensed bosons as being in the ground energy state,
and the uncondensed bosons as being in excited energy states.
The present author instead takes the condensed bosons to be
those in multiply occupied low-lying momentum states
(Attard 2025a),
although this is not relevant for the present paper.

In the above $m$ is the atomic mass of $^4$He,
${\bf v}$ is the velocity, $n=n_0+n_*$ is the total number density,
$p$ is the pressure, $T$ is the temperature,
and $\sigma$ is the entropy density.
These are functions of position ${\bf r}$ and time $t$.
These equations have been linearized in velocity,
keeping the shear viscosity $\eta$
(also called the dynamic viscosity).

These equations differ from the usual equations of classical hydrodynamics
by the inclusion of the entropy term.
This is necessary for consonance with the fountain pressure equations,
as we shall see.
Also, there is no contribution from the viscosity
to retard the superfluid flow.

The two-fluid  equations
are often combined with an expression
for the rate of change of entropy density,
\begin{equation} \label{Eq:dot-sigma}
\frac{\partial \sigma}{\partial t}
=
- \nabla \cdot(\sigma {\bf v}_*).
\end{equation}
This is a serious statement about entropy transport in the superfluid regime.
Many interpret it to mean both that entropy is conserved,
and that condensed bosons carry no entropy.
The present author disputes both assertions.

Nevertheless from this and the two-fluid model equations
the wave equations for first and second sound may be derived
(Landau 1941, Tisza 1938).
Add together the two equations and neglect the viscosity,
\begin{equation}
m n_0 \frac{\partial {\bf v}_0}{\partial t}
+
m n_* \frac{\partial {\bf v}_*}{\partial t}
= - \nabla p.
\end{equation}
Notice that it is necessary for the entropy terms to have opposite
sign and the same density prefactor in order to cancel here,
which is one way of checking for misprints in the two equations.
Denote the time derivative by a dot and take the gradient of this
\begin{equation}
m \nabla \cdot ( n_0 \dot {\bf v}_0)
+
m \nabla \cdot ( n_* \dot {\bf v}_*)
= - \nabla^2 p .
\end{equation}
The rate of change of density is the divergence of the number flux,
$\dot n_k = - \nabla \cdot {\bf J}_{{\rm N},k} = -\nabla\cdot(n_k{\bf v}_k)$,
which gives the second time derivative as
\begin{equation}
\frac{\partial^2 n_k}{\partial t^2}
=
- \nabla \cdot (  n_k \dot{\bf v}_k)
- \nabla \cdot ( \dot n_k {\bf v}_k) ,
\;\; k\in\{0,*\} .
\end{equation}
The final term is second order in velocity and may be neglected.
Adding together what remains gives
\begin{equation}
m\frac{\partial^2 n}{\partial t^2}
=
\nabla^2 p .
\end{equation}
This is the equation for first sound,
which gives oscillating pressure and density waves.

For second sound,
take the time derivative of Eq.~(\ref{Eq:dot-sigma}),
use Eq.~(\ref{Eq:TwoFluid*}),
and neglect second order velocity terms
and all the gradients except that of temperature,
\begin{eqnarray}
\frac{\partial^2 \sigma}{\partial t^2}
& = &
\frac{-\partial }{\partial t}
\left\{  {\bf v}_* \cdot \nabla \sigma
+ \sigma \nabla \cdot {\bf v}_* \right\}
\nonumber \\ & \approx &
- \sigma \nabla \cdot
\left[
\frac {-1}{m n }\nabla p
-\frac{n_0}{mnn_*}\sigma \nabla T
\right]
\nonumber \\ & \approx &
\frac{n_0\sigma^2 }{mnn_*}\nabla^2 T .
\end{eqnarray}
This is the equation for second sound,
which gives oscillating temperature and entropy waves.

The existence of second sound
has been confirmed experimentally
(see the review by Donnelly (2009)).
This confirms the validity of the two-fluid theory,
Eqs~(\ref{Eq:TwoFluid0}), (\ref{Eq:TwoFluid*}), and (\ref{Eq:dot-sigma}).
Nevertheless
there remain significant questions regarding their interpretation
and the nature of the approximations that underpin them.
One such question is whether the dissipation rate,
Eq.~(\ref{Eq:dot-sigma}),
is an independent  axiom in two-fluid theory,
or whether it is a consequence of
Eqs~(\ref{Eq:TwoFluid0}) and (\ref{Eq:TwoFluid*}).

The point of the present paper
is to derive these equations from first principles.
It is hoped that in so doing the physical nature of condensation
and its role in superfluid flow will be clarified,
and the regime of applicability of the approximations that are invoked
will be delineated.

\subsection{Empirical Equations from Fountain Pressure Measurements}

The present derivation takes as axiomatic
certain principles that have been established
from fountain pressure measurements of steady superfluid flow.
Specifically the fountain pressure equation
for the rate of change of pressure
of a higher temperature chamber $B$
connected to a chamber $A$ by steady superfluid flow is
(H. London 1939)
\begin{equation} \label{Eq:HLondon}
\frac{\mathrm{d}p_B(T_B|T_A,p_A)}{\mathrm{d}T_B} = \sigma_B.
\end{equation}
This fits measured data, apparently exactly.

The thermodynamic implication of this equation is
that regions connected by steady superfluid flow
have equal chemical potential
(Attard 2025a Ch.~4),
\begin{equation} \label{Eq:muA=muB}
\mu_A = \mu_B.
\end{equation}
This in turn implies that condensed bosons minimize the energy
at constant entropy,
$\partial E(S,V,N)/\partial N = \mu$,
the chemical potential $\mu$ being the mechanical (ie.\ non-entropic)
part of the energy.
Thus the convective energy flux for condensed bosons is
\begin{equation} \label{Eq:JE0conv}
{\bf J}_{{\rm E},0}^{\rm conv}
=
\mu n_0 {\bf v}_0 .
\end{equation}
The fountain pressure equation
says that condensed bosons move steadily with constant entropy,
not with zero entropy,
which is consistent with the general requirements
of equilibrium statistical mechanics (Attard 2025a Ch.~5,  2025b).

From this it follows that the acceleration
is the negative gradient of the chemical potential,
\begin{equation} \label{Eq:dotv0}
m \frac{\partial {\bf v}_0}{\partial t} = - \nabla \mu.
\end{equation}
The reason that this
is not the gradient in average energy
is that condensed bosons move at constant entropy.

On the one hand these two equations
are empirical equations based on fountain pressure measurements
of superfluid flow.
On the other hand they can be derived from the general requirements
of equilibrium statistical mechanics (Attard 2025a Ch.~5, Attard 2025b).

The Gibbs-Duhem equation (\ref{Eq:GD}) gives this as
\begin{eqnarray} \label{Eq:dotv0'}
m n_0 \frac{\partial {\bf v}_0}{\partial t}
& = &
- n_0 \nabla \mu
\nonumber \\ & = &
\frac{n_0}{n} \sigma \nabla T - \frac{n_0}{n} \nabla p
- \frac{n_0^2}{n} \nabla\!\left( \psi + \frac{m v_0^2}{2} \right)
\nonumber \\ & & \mbox{ }
- \frac{n_0n_*}{n}  \nabla\!\left( \psi + \frac{m v_*^2}{2} \right) .
\end{eqnarray}
This is the same as Eq.~(\ref{Eq:TwoFluid0}),
apart from the external potential and second order velocity contributions.

%
\section{Derivation}
\setcounter{equation}{0} \setcounter{subsubsection}{0}
\renewcommand{\theequation}{\arabic{section}.\arabic{equation}}
%

Hydrodynamics is based on conservation laws and fluxes,
and also thermodynamic relationships.
Fundamentally the following analysis
comes from  de Groot and Mazur (1984),
with certain thermodynamic results from Attard (2002).
The rate of change of entropy density
yields the general equations of fluctuating hydrodynamics
(Attard 2012 Ch.~5),
with suitable modifications for the superfluid regime.

\subsubsection{A Little Thermodynamics} \label{Sec:TD}

Consider two systems at the same density
with and without a uniform external potential $\psi$.
If isolated, the entropy is the same if the bare system has energy
$E' = E - N\psi$.
That is, $S(E,N,V;\psi) = S(E',N,V)$.
Evidently the temperatures are also equal,
$T^{-1} = \partial S(E,N,V;\psi)/\partial E
= \partial S(E',N,V)/\partial E'$,
since the two derivatives are the same at constant $\psi$.
The number derivative at constant $E$ is
$-\mu/T = \partial S(E,N,V;\psi)/\partial N$.
Hence $\partial S(E-N\psi,N,V)/\partial N = [-\psi-\mu']/T $,
or $\mu' = \mu - \psi$.
In a grand canonical system,
$\mu'$ is the chemical potential for the bare system
that would give the same density as the actual system
with the external potential.
The volume derivative is unchanged, $p'=p$.

In terms of the energy density $\varepsilon=E/V$
and number density $n=N/V$,
the entropy density is
$\sigma(\varepsilon,n;\psi) = \sigma(\varepsilon',n)$,
with $\varepsilon' = \varepsilon - n \psi$.
The Gibbs equation (derived for a grand canonical system)
for the full and bare systems is
\begin{eqnarray}
\sigma & = &
\frac{1}{T} [ \varepsilon + p - n \mu ]
\nonumber \\ & = &
\frac{1}{T}  [ \varepsilon' + p - n \mu' ].
\end{eqnarray}
That these are equal
says that a slowly varying external potential
has no non-trivial effect on the molecular configurations,
since these are determined by the bare part of the energy.

Formally, the total change of the entropy density is
\begin{eqnarray}
\Delta\sigma
 & = &
\Delta (\varepsilon/T) 
+ \Delta (p/T) - \Delta(n\mu/T)
\nonumber \\ & = &
\Delta (\varepsilon'/T)
+ \Delta (p/T) - \Delta(n\mu'/T).
\end{eqnarray}
But we also have the standard thermodynamic definition
$\Delta \sigma = T^{-1} \Delta \varepsilon'   - (\mu'/T) \Delta n$.
The Gibbs-Duhem equation can be derived by subtracting these,
\begin{eqnarray}
0 & = &
\varepsilon' \Delta T^{-1}  + \Delta(p/T) - n \Delta(\mu'/T)
 \\ & = &
T \sigma \Delta T^{-1} +  T^{-1} \Delta p - (n/T) \Delta \mu'
\nonumber \\ & = &
T \sigma \Delta T^{-1} +  T^{-1} \Delta p
- (n/T) \Delta \mu  + (n/T) \Delta \psi, \nonumber
\end{eqnarray}
or $0 = \sigma \Delta T - \Delta p + n  \Delta \mu  - n \Delta \psi$.

In the present problem of superfluid hydrodynamics,
we have two species,
condensed bosons with number density $n_0$ and momentum density ${\bf p}_0$,
and uncondensed bosons
with number density $n_*$ and momentum density ${\bf p}_*$.
The total number density is $n = n_0 + n_*$.
The kinetic energy associated with this average motion
is slowly varying in space and time,
and is treated in the same way as the external potential.
We define the bare (or internal) energy density
\begin{equation} 
\varepsilon^{(0)}
\equiv
\varepsilon
- n_0 \psi - \frac{p_0^2}{2mn_0}
- n_* \psi - \frac{p_*^2}{2mn_*} .
\end{equation}

Because the bosons can change their state
in the actual system, $0 \Leftrightarrow *$,
we must have $\mu_0 = \mu_* \equiv \mu$.
We therefore define the bare chemical potentials to be
\begin{equation} 
\mu^{(0)}_0
\equiv
\mu - \psi - \frac{p_0^2}{2mn_0^2}
=
\mu - \psi - \frac{m v_0^2}{2}
,
\end{equation}
and
\begin{equation} 
\mu^{(0)}_*
\equiv
\mu - \psi - \frac{p_*^2}{2mn_*^2}
=
\mu - \psi - \frac{m v_*^2}{2} .
\end{equation}
The Gibbs equation is
\begin{eqnarray}
\sigma = \sigma^{(0)}
& = &
\frac{1}{T}
[\varepsilon + p - n_0 \mu  - n_* \mu  ]
\nonumber \\ & = &
\frac{1}{T}
[\varepsilon^{(0)} + p - n_0 \mu^{(0)}_0  - n_* \mu^{(0)}_*  ].
\end{eqnarray}

Formally, the total change of the entropy density is
\begin{eqnarray}
\Delta\sigma
 & = &
\Delta (\varepsilon/T) 
+ \Delta (p/T) - \Delta(n\mu/T)
 \\ & = & \nonumber
\Delta \frac{\varepsilon^{(0)}}{T}
+ \Delta \frac{p}{T}
- \Delta \frac{n_0 \mu^{(0)}_0}{T} - \Delta \frac{n_* \mu^{(0)}_*}{T}.
\end{eqnarray}
But we also have the standard thermodynamic definition,
$\Delta \sigma = T^{-1} \Delta \varepsilon^{(0)}
 - (\mu^{(0)}_0/T) \Delta n_0  - (\mu^{(0)}_*/T) \Delta n_*$.
The Gibbs-Duhem equation comes from subtracting these
\begin{eqnarray}
0 & = &
\varepsilon^{(0)} \Delta \frac{1}{T}
+ \Delta \frac{p}{T}
- n_0 \Delta\frac{\mu^{(0)}_0}{T}
- n_* \Delta\frac{\mu^{(0)}_*}{T}
\nonumber \\ & = &
T \sigma \Delta \frac{1}{T} +  \frac{1}{T}  \Delta p
- \frac{n_0}{T} \Delta \mu^{(0)}_0
- \frac{n_*}{T} \Delta \mu^{(0)}_*
\nonumber \\ & = &
T \sigma \Delta\frac{1}{T} +  \frac{1}{T} \Delta p - \frac{n}{T} \Delta \mu
\nonumber \\ & & \mbox{ }
+ \frac{n_0}{T} \Delta\!\left( \psi + \frac{m v_0^2}{2} \right)
+ \frac{n_*}{T} \Delta\!\left( \psi + \frac{m v_*^2}{2} \right)\!,
\end{eqnarray}
or
\begin{eqnarray} \label{Eq:GD}
\sigma \Delta T
& = &
\Delta p - n_0 \Delta \mu_0^{(0)} - n_* \Delta \mu_*^{(0)}
 \\ & = &
\Delta p - n  \Delta \mu
\nonumber \\ & & \mbox{ }\nonumber
+ n_0 \Delta\!\left( \psi + \frac{m v_0^2}{2} \right)
+ n_* \Delta\!\left( \psi + \frac{m v_*^2}{2} \right).
\end{eqnarray}

\subsubsection{Fluxes}

We have two species, condensed and uncondensed bosons,
denoted $k=0$ and $k=*$, respectively.
These bosons can change their state, $0 \Leftrightarrow *$,
with the rate of change of number density due to this reaction being
\begin{equation}
\dot n_{0}^\mathrm{react}
= - \dot n_{*}^\mathrm{react} .
\end{equation}
A positive value means nett creation.
With this, the rate of change of number density is
\begin{equation}
\frac{\partial n_k }{\partial t}
=
\dot n_{k}^\mathrm{react} - \nabla \cdot {\bf J}_{{\rm N},k} ,
\end{equation}
with the number flux being
${\bf J}_{{\rm N},k}  = n_k  {\bf v}_k$,
${\bf v}_k$ being the velocity of species $k$.

The energy flux consists of the diffusive flux,
${\bf J}_{\rm E}^0$, otherwise known as the heat flow,
the convective flux ${\bf J}_{{\rm E},k}^{\rm conv}$,
and energy flux due to work done, ${\bf J}_{\rm E}^{\rm work}$.
The convective energy flux for the condensed bosons is
given by Eq.~(\ref{Eq:JE0conv}) based on the fountain pressure equation,
\begin{equation}
{\bf J}_{{\rm E},0}^{\rm conv}
=
\mu n_0  {\bf v}_0  .
\end{equation}
This says that only the mechanical part of the energy,
$\mu = \partial E(S,V,N;\psi)/\partial N$,
is transported by the condensed bosons.
This is why the total rather than the bare chemical potential appears.
In view of this, the convective energy flux for the uncondensed bosons is
\begin{equation}
{\bf J}_{{\rm E},*}^{\rm conv}
=
[\varepsilon - n_0  \mu ]   {\bf v}_*  .
\end{equation}
We have to remove the mechanical energy
of the condensed bosons from the total energy density
because this is not carried by the uncondensed bosons.
The energy flux due to the work done
on the hydrodynamic volume carried by ${\bf v}_*$ is
\begin{equation}
{\bf J}_{{\rm E}}^{\rm work}
=  \underline{ \underline P}  \cdot {\bf v}_* .
\end{equation}

With these the rate of change of the energy density is
\begin{equation}
\frac{\partial \varepsilon  }{\partial t}
=
n \frac{\partial \psi }{\partial t}
- \nabla \cdot {\bf J}_{\rm E},
\end{equation}
where the total number density is
$n({\bf r},t) = n_0({\bf r},t) + n_*({\bf r},t)$.
The energy flux is
\begin{eqnarray}
{\bf J}_{{\rm E}}
& = &
{\bf J}_{{\rm E}}^0
+ {\bf J}_{{\rm E},0}^{\rm conv} + {\bf J}_{{\rm E},*}^{\rm conv}
+ {\bf J}_{{\rm E}}^\mathrm{work}
\nonumber \\ & = &
{\bf J}_{{\rm E}}^0
+ \mu n_0 {\bf v}_0
+ [\varepsilon - n_0  \mu  ]  {\bf v}_*
+ \underline{ \underline P} \cdot {\bf v}_* .
\end{eqnarray}

The momentum density is
${\bf p}_k({\bf r},t) = m n_k({\bf r},t) {\bf v}_k({\bf r},t)$.
The fountain pressure result for the acceleration of a condensed boson,
Eq.~(\ref{Eq:dotv0}),
is a source for the rate of change of the  momentum density.
Subtracting the divergence of the momentum flux gives
\begin{equation} \label{Eq:dotp0}
\frac{\partial {\bf p}_0}{\partial t}
=
- n_0 \nabla \mu - \nabla \cdot ({\bf p}_0 {\bf v}_0 ).
\end{equation}

Arguably the reaction rate is also a source of momentum,
such as $m \dot n_{0}^\mathrm{react} {\bf v}_0$.
Perhaps ${\bf v}_0$ could be replaced by ${\bf v}_*$
for $ \dot n_{0}^\mathrm{react} >0$.
Since there is some uncertainty in this term,
and since it is second order in velocity
in the rate of entropy production below,
we do not include it.

This can be rewritten to give the rate of change
of the condensed bosons' velocity,
\begin{eqnarray} \label{Eq:dotp0''}
m n_0 \frac{\partial {\bf v}_0}{\partial t}
& = &
-m {\bf v}_0 \frac{\partial n_0}{\partial t}
- n_0 \nabla \mu
- \nabla \cdot ( m n_0 {\bf v}_0 {\bf v}_0 )
\nonumber \\ & = &
m {\bf v}_0 \nabla \cdot(n_0{\bf v}_0)
- n_0 \nabla \mu
- \nabla \cdot ( m n_0 {\bf v}_0 {\bf v}_0 )
\nonumber \\ & = &
- n_0 \nabla \mu
- m n_0 {\bf v}_0 \cdot  \nabla  {\bf v}_0.  
\end{eqnarray}
(The final term in the final equality
has been corrected from v2.)
This  includes the non-linear velocity term
and is a more exact version of the two-fluid
Eqs~(\ref{Eq:TwoFluid0}) and (\ref{Eq:dotv0'}).

The rate of change of momentum density for the uncondensed bosons is
\begin{equation} \label{Eq:dotp*}
\frac{\partial{\bf p}_*}{\partial t}
=
n_0 \nabla \mu
-  n \nabla \psi    
- \nabla \cdot \underline{ \underline P}
- \nabla \cdot [ {\bf p}_* {\bf v}_* ].
\end{equation}
The first term
is the negative of the source part of  ${\partial {\bf p}_0}/{\partial t}$,
and the next two terms are the total change in momentum due to forces.
The difference in these
is the forced change in momentum of the uncondensed bosons alone.
The pressure tensor appears
and the contribution from the external potential are explicit.
The remaining term is the divergence of the momentum flux.
This is a non-linear version of the two-fluid theory,
Eq.~(\ref{Eq:TwoFluid*}),
the pressure tensor being replaced as usual by the pressure
and the viscous pressure tensor,
$\underline{ \underline P}
= [p + \pi ]\underline{ \underline {\rm I}}
+ \underline{ \underline \Pi}^*$,
with the most likely traceless viscous pressure tensor
being proportional to the shear rate tensor,
$\overline{\underline{ \underline \Pi}}^* =
- 2 \eta [\nabla {\bf v}_*]^{*,{\rm sym}}$
(Attard 2012 \S 5.3).

\subsubsection{Entropy Production}

The various thermodynamic derivatives of the entropy
can be taken as usual
(Appendix~\ref{Sec:ds/de-etc}).
These and the above results give the rate of change of entropy density as
\begin{eqnarray}
\frac{\partial \sigma}{\partial t}
& = &
\frac{1}{T} \frac{\partial \varepsilon}{\partial t}
-
\frac{n}{T} \frac{\partial \psi}{\partial t}
- \frac{{\bf v}_0}{T} \cdot \frac{\partial {\bf p}_0}{\partial t}
- \frac{{\bf v}_*}{T} \cdot \frac{\partial {\bf p}_*}{\partial t}
\nonumber \\ && \mbox{ }
- \frac{\mu - v_0^2/m}{T} \frac{\partial n_0}{\partial t}
- \frac{\mu - v_*^2/m}{T} \frac{\partial n_*}{\partial t}
\nonumber \\ & = & 
- \frac{1}{T}  \nabla \!\cdot\!
\left\{ {\bf J}_{{\rm E}}^0
+ \mu n_0 {\bf v}_0
+ [\varepsilon - n_0  \mu  ]  {\bf v}_*
+ \underline{ \underline P}\cdot {\bf v}_* \right\}
\nonumber \\ && \mbox{ }
- \frac{{\bf v}_0}{T} \cdot
\left\{ - n_0 \nabla \mu - \nabla \cdot ({\bf p}_0 {\bf v}_0 ) \right\}
\nonumber \\ && \mbox{ }
- \frac{{\bf v}_*}{T} \cdot
\left\{ n_0 \nabla \mu
-  n \nabla \psi
- \nabla \cdot \underline{ \underline P}
- \nabla \cdot ( {\bf p}_* {\bf v}_* ) \right\}
\nonumber \\ && \mbox{ }
- \frac{\mu - v_0^2/m}{T}
\left\{ \dot n_0^\mathrm{react} - \nabla \cdot (n_0 {\bf v}_0) \right\}
\nonumber \\ && \mbox{ }
- \frac{\mu - v_*^2/m}{T}
\left\{ \dot n_*^\mathrm{react} - \nabla \cdot (n_* {\bf v}_*) \right\} .
\end{eqnarray}
The terms involving gradients of $\mu n_0 {\bf v}_0$ cancel.
The terms involving the pressure tensor are
\begin{eqnarray}
\mbox{RHS}_{\rm P}
& = &
\frac{-1}{T} \nabla \cdot (\underline{ \underline P}  \cdot {\bf v}_*)
+ \frac{1}{  T}{\bf v}_* \nabla : \underline{ \underline P}
\nonumber \\ & = &
\frac{-1}{T} \underline{ \underline P} : \nabla {\bf v}_*
\nonumber \\ & = &
\frac{-p}{T} \nabla \cdot {\bf v}_*
- \frac{1}{T}  \underline{ \underline \Pi} : \nabla {\bf v}_* .
\end{eqnarray}
The most likely traceless viscous pressure tensor
is  proportional to the shear rate tensor,
$\overline{\underline{ \underline \Pi}}^* =
- 2 \eta [\nabla {\bf v}_*]^{*,{\rm sym}}$,
and so the final term is higher order in the velocity
(Attard 2012 \S 5.3).
Dropping the heat flux,
the viscous pressure tensor,
and the explicitly higher order velocity terms
gives
\begin{eqnarray}
\frac{\partial \sigma}{\partial t}
 & \approx &
\frac{ - 1}{T}  \nabla \cdot
\left\{ [\varepsilon - n_0  \mu  ]  {\bf v}_* \right\}
- \frac{{\bf v}_*}{T} \cdot
\left\{ n_0 \nabla \mu -  n \nabla \psi \right\}
\nonumber \\ && \mbox{ }
- \frac{p}{T} \nabla \cdot {\bf v}^*
+ \frac{\mu}{T}\nabla \cdot (n_* {\bf v}_*)
\nonumber \\ & = & 
\left\{ \frac{-\varepsilon}{T} - \frac{p}{T} + \frac{n \mu}{T} \right\}
\nabla \cdot {\bf v}^*
\nonumber \\ && \mbox{ }
-\frac{ {\bf v}_*}{T} \cdot
\left\{ n_0 \nabla \mu - n \nabla \psi
+ \nabla [\varepsilon - n_0  \mu ]
- \mu \nabla n_* \right\}
\nonumber \\ & = & 
-\sigma \nabla \cdot {\bf v}^*
-\frac{ {\bf v}_*}{T} \cdot
\left\{ -n \nabla \psi + \nabla \varepsilon  - \mu \nabla n \right\}
\nonumber \\ & = & 
-\sigma \nabla \cdot {\bf v}^*
-\frac{ {\bf v}_*}{T} \cdot
\left\{  \nabla \varepsilon^{(0)} - \mu^{(0)} \nabla n  \right\}
\nonumber \\ & = & 
-\nabla \cdot(\sigma {\bf v}_*) .
\end{eqnarray}
This is just the two-fluid theory equation
for the rate of change of entropy density,
Eq.~(\ref{Eq:dot-sigma}).
This shows that
Eq.~(\ref{Eq:dot-sigma}) is a consequence of
Eqs~(\ref{Eq:TwoFluid0}) and (\ref{Eq:TwoFluid*})
rather than an independent axiom or approximation.
This result for the rate of change of entropy density
holds in the linear velocity regime,
and where the heat flux is negligible compared to convection.

It should be noted that in deriving this result we have not assumed
that the entropy of condensed bosons is zero.
We have used the experimental and theoretical fact
that condensed bosons minimize the energy at constant entropy,
Eq.~(\ref{Eq:JE0conv}) (Attard 2025a Ch.~5,  2025b).

%
\section{Fountain Pressure Paradox}
\setcounter{equation}{0} \setcounter{subsubsection}{0}
\renewcommand{\theequation}{\arabic{section}.\arabic{equation}}
%

There is a paradox at the heart of the fountain pressure.
On the one hand H. London's equation  (\ref{Eq:HLondon})
implies equal chemical potential, $\mu_A=\mu_B$,
which fits the experimental data with great accuracy.
On the other hand,
\emph{if} the chemical potentials were exactly equal
\emph{then} there would be nothing to distinguish the two chambers
and nothing to determine the direction of superfluid flow.
We can resolve this paradox quantitatively as follows.

The low temperature, saturated chamber $A$ is located at $x=0$
and the high temperature, high pressure chamber $B$ is at $x=L >0$.
The closed chambers are connected by a capillary of radius $R \ll L$
that is long enough to ignore quantization of the longitudinal momentum.
There is  Poiseuille flow of uncondensed bosons from $B$ to $A$,
and superfluid flow of condensed bosons in the opposite direction.

We use the Gibbs-Duhem equation~(\ref{Eq:GD})
to write the two-fluid equation (\ref{Eq:TwoFluid*})
for the uncondensed bosons as
\begin{eqnarray}
m n_* \frac {\partial {\bf v}_*}{\partial t}
& = &
\frac {-n_*}{n} \nabla p
- \frac{n_0}{n}  [ \nabla p - n \nabla \mu ]
+ \eta \nabla^2 {\bf v}_*
\nonumber \\ & \approx &
- \nabla p
+ \eta \nabla^2 {\bf v}_* ,
\end{eqnarray}
the  gradient of the chemical potential being negligible.
From this  the usual Navier-Stokes equation
gives the rate of change of the number of uncondensed bosons
in the high temperature chamber as
(Wikipedia 2025)
\begin{equation}
\dot N_{*}(L) = -n_*(L) \frac{\pi R^4 p_{BA}}{8 \eta L} .
\end{equation}

In contrast,
the rate of change of the number of condensed bosons in the same chamber is
\begin{equation}
\dot N_{0}(L) = n_0(L) \pi R^2  v_{0,x}(L) .
\end{equation}
In the steady state these must be equal and opposite.
Accepting that the number densities and the viscosity
can be measured experimentally (Donnelly and  Barenghi 1998)
means that these equations determine $v_{0,x}(L)$
in terms of the fountain pressure difference $p_{BA} = p_B -p_A$
and the length and radius of the cylindrical capillary.

Here $v_{0,x}(L)>0$ is the average velocity of the condensed bosons
at the end of the capillary where it enters the high temperature chamber $B$.
We assume that at the beginning of the capillary,
where the bosons leave the low temperature chamber $A$,
the average velocity of the condensed bosons is zero, $v_{0,x}(0) = 0$.

The fountain pressure equation
gives the acceleration of condensed bosons,
Eqs~(\ref{Eq:TwoFluid0}) or (\ref{Eq:dotv0}),
\begin{equation} 
\frac{\partial {v}_{0,x}}{\partial t}
= \frac{ -1}{m} \frac{ \partial \mu}{\partial x}
\equiv a.
\end{equation}
We shall take the acceleration $a$ to be constant.
That this is positive follows because
with $T_B > T_A$,
$ \mu_B^{\rm sat} < \mu_A^{\rm sat}$  (Attard 2025a \S 4.6),
and so the chemical potential of chamber $B$
may increase toward that of chamber $A$ without ever quite reaching it.
For constant acceleration,
the transit time for a condensed boson through the capillary is
$t_L = \sqrt{2L/a}$,
and the final velocity is
$v_{0,x}(L) = a t_L = \sqrt{2La}$.
Hence the acceleration is
\begin{eqnarray}
\frac{\mu_A-\mu_B}{mL}
\equiv a
& = & \frac{v_{0,x}(L)^2}{2L}
\nonumber \\ & = &
\frac{1}{2L}
\left[ \frac{ n_*(L) R^2 p_{BA}  }{8  n_0(L) \eta L } \right]^2 .
\end{eqnarray}
This gives the difference in chemical potential as
\begin{equation}
\mu_A-\mu_B
=
\frac{m}{2}
\left[ \frac{ n_*(L) R p_{BA}  }{8  n_0(L) \eta } \right]^2
\frac{R^2}{L^2} .
\end{equation}
Experimental tests would be challenging because
typical fountain pressure measurements  have
slit width to length ratio $W/L = {\cal O}(10^{-6})$
(Hammel and Keller 1961).
Nevertheless conceptually this resolves the paradox outlined above:
the difference in the chemical potentials is small enough to escape
experimental detection,
but sufficient to determine the direction of superfluid flow.

%
\section{Revolutionary Analysis of Rotational Motion} \label{Sec:Revolve}
\setcounter{equation}{0} \setcounter{subsubsection}{0}
\renewcommand{\theequation}{\arabic{section}.\arabic{equation}}
%

\subsection{Background}

Landau (1941) asserted that superfluid flow
was irrotational, $\nabla \times {\bf v}_0 = {\bf 0}$
(ie.\ the vorticity vanishes everywhere),
and this has since been taken as a fundamental principle.
The notion is apparently motivated by two observations:
First,  inviscid classical fluids are irrotational,
and superfluids certainly lack viscosity.
Second,
a superconducting current is irrotational,
as shown by
the (second) London equation for superconductivity
(F and H London 1935),
which accounts for the Meissner effect
by which a magnetic field is excluded from a superconductor.

The theoretical justification for irrotational superfluid flow
is limited (see \S\S \ref{Sec:nonVortex},
\ref{Sec:Landau}, \ref{Sec:QuantCirc}, and \ref{Sec:MacroWave}).
Possibly  Landau's principle might be based on a false analogy.
For example,
inviscid classical fluids are an idealization
that might possibly apply at gas-like densities,
whereas helium~II is a liquid.
Or the Meissner effect and the London equation for superconducting currents
are a direct consequence of Maxwell's equations,
and it is a curly question as to whether
these apply to superfluidity in helium~II,
which has no electrical currents.
Nor is there any unambiguous experimental evidence to support Landau's idea.
In fact, as will be shown shortly,
the experimental evidence ostensibly refutes the notion.
Unfortunately rather than accept the obvious conclusion that Landau's (1941)
theory is untenable,
workers have twisted the flow so as to preserve Landau's idea
(see below).

The issue of whether or not superfluid flow is irrotational
(ie.\ has zero vorticity)
is important for several reasons.
If true,
then the existence of a velocity potential restricts and simplifies
the allowed solutions to the two-fluid model hydrodynamic equations.
Another reason is that it gives insight into the behavior and understanding
of superfluidity at the molecular level.
In particular Landau's (1941) theory for the origin of superfluidity,
for which he was awarded the Nobel prize (Physics 1962),
introduces rotons
as a type of rotational first excited state
whose occupancy is taken to be inconsistent with superfluidity.
Onsager (1949) (Nobel Laureate in Chemistry 1968)
suggested that
`Vortices in a suprafluid are presumably quantized;
the quantum of circulation is $h/m$'.
Feynman (1955) (Nobel Laureate in Physics 1965)
`was the first to suggest that the formation of vortices in liquid helium II
might provide the mechanism responsible
for the breakdown of superfluidity in the liquid'
(Pathria 1972 \S10.8).
This idea has been further developed,
successively improving agreement with measured data
(Fetter 1963, Kawatra and Pathria 1966, Pathria 1972 \S10.8).
(Some improvement was needed.
Landau's (1941) Nobel prize winning roton formulation
originally overestimated  the critical velocity
by several orders of magnitude.)

Rotational motion in helium~II,
specifically the damping of a torsional pendulum,
was one of the original methods for measuring its viscosity.
Oscillatory and steady rotation are qualitatively different
for the superfluid (see below),
with the latter being the focus here.

\subsection{Steady Rotation}

Osborne (1950) measured the steady rotation of a bucket of He~II
and established that the free surface of the liquid
had the classic  parabolic shape,
\begin{equation}
z_{\rm surf}({\bf r}) = z_0 + \frac{\omega^2 r^2}{2g},
\end{equation}
where $\omega$ is the angular velocity,
$r$ is the radius from the $z$ axis,
and $g$ is the acceleration due to gravity.
This implies that the whole liquid is rotating rigidly.
If Landau's (1941) irrotational idea held,
only the normal liquid can rotate
and the quadratic term should be scaled by
the fraction of normal liquid, $n_*/n$.

Instead of drawing the obvious conclusion that the experimental data
disproved Landau's (1941) hypothesis,
workers sought refuge in complications.
Walmsley and Lane (1958) (see also Lane (1962) and Pathria (1972 \S10.7))
attribute to Landau and Lifshitz (1955) and to Lifshitz and Kagenov (1955)
a model of microscopic vortices of superfluid,
with velocity inversely proportional
to the distance from their axis and zero vorticity,
arranged in annular rings in such a way as to give
the same overall rotation velocity as the normal fluid (Pathria 1972 \S10.7)
(see \S\ref{Sec:nonVortex}).
Walmsley and Lane (1958) state
that the trial wave functions of Onsager and Feynman
give `an equilibrium distribution of vortices [that] leads
to an angular momentum and free energy which are
indistinguishable from those of a classical liquid\ldots'
and
`that this model compares favorably
with that of Landau and Lifshitz'.

In any case, although the individual vortices have zero vorticity
(but see \S \ref{Sec:nonVortex}),
the overall superfluid velocity field corresponding to
their  uniform distribution is one of rigid rotation
with uniform vorticity,
$\nabla \times {\bf v}_0({\bf r}) = 2\omega \hat{\bf z}$
(Pathria 1972 \S10.7).
It is not clear to the present author
how this can be considered to be irrotational ( \S\ref{Sec:nonVortex}).

A simpler explanation of Osborne's (1950) measurements
is given by the two-fluid theory.
The velocity equation (\ref{Eq:dotp0''}) for condensed bosons is
\begin{equation}
m n_0 \frac{\partial {\bf v}_0}{\partial t}
=
- n_0 \nabla \mu
- n_0 m {\bf v}_0 \cdot  \nabla  {\bf v}_0.  
\end{equation}
For uncondensed bosons (cf.\ Eq.~(\ref{Eq:dotp*})) it is
\begin{eqnarray}
\lefteqn{
m n_* \frac{\partial {\bf v}_*}{\partial t}
}  \\
& = &
n_0 \nabla \mu
-  n \nabla \psi
- \nabla \cdot \underline{ \underline P}
- n_* m {\bf v}_*  \cdot \nabla {\bf v}_*
\nonumber \\ & \approx &\nonumber
n_0 \nabla \mu
-  n \nabla \psi
- \nabla p
+ \eta \nabla^2 {\bf v}_*
- n_* m {\bf v}_* \cdot \nabla {\bf v}_* .
\end{eqnarray}

We are interested in a cylindrical system
steadily rotating about the $z$-axis with angular velocity $\omega$.
For the present mixture of superfluid and normal fluid,
in the steady state,
${\partial {\bf v}_0}/{\partial t} = {\bf 0}$,
the condensed boson velocity field satisfies
\begin{equation}
n_0 \nabla \mu
=
- n_0 m  {\bf v}_0 \cdot  \nabla  {\bf v}_0.  
\end{equation}
We insert this into the equation
for the rate of change of the velocity of the uncondensed bosons.
In the steady state and neglecting the shear viscosity this yields
\begin{eqnarray}
{\bf 0}
& = &
-  \nabla \psi
- \nabla p
- n_* m {\bf v}_* \cdot \nabla {\bf v}_*
- n_0 m {\bf v}_0 \cdot  \nabla  {\bf v}_0
\nonumber \\ & = &
-  \nabla \psi - \nabla p
- n m {\bf v}_* \cdot \nabla {\bf v}_* ,
\end{eqnarray}
where the total number density is $n({\bf r}) = n_0({\bf r})+n_*({\bf r})$.
In the second equality we have taken
the superfluid and normal fluid velocity fields
to be equal,  ${\bf v}_0({\bf r}) = {\bf v}_*({\bf r})$.
One can confirm by direct substitution
that for the gravitational potential, $\psi = m g z$,
and rigid rotation,
${\bf v}_0({\bf r})  = {\bf v}_*({\bf r}) =   \omega r \hat{\bm \theta}$,
the classical pressure profile satisfies this,
$p({\bf r}) = p_0 - n mg z + n m \omega^2 r^2/2$.

The pressure is atmospheric at the surface,
$p(r,\theta,z_{\rm surf}(r)) = p(0,\theta,z_{\rm surf}(0))
=p_0 - n mg z_0$,
which gives the free surface profile as
$z_{\rm surf}(r) = z_0 + \omega^2 r^2/2g$.
This is the result measured by Osborne (1950)
for $^4$He below the $\lambda$-transition.

The above equations for the rate of change of each velocity  neglect
the contribution from the reaction rate,
$\dot n_0^{\rm react}({\bf r}) = -\dot n_*^{\rm react}({\bf r})$.
This likely provides the molecular mechanism by which
the velocity profile of the condensed bosons becomes established.

Obviously the superfluid is \emph{not} irrotational,
\begin{equation}
\nabla \times {\bf v}_0({\bf r})
=
2 \omega \hat{\bf z} ,
\end{equation}
which  contradicts Landau's principle.

One might object that the analysis only shows
that the two-fluid theory gives a possible solution to the problem.
It could be argued that the Landau (1941) irrotational condition,
$\nabla \times {\bf v}_0({\bf r}) ={\bf 0}$,
is an independent  requirement that should be added to the two-fluid theory,
which would rule out the present solution.

To answer this we can appeal to the physical plausibility of the result
and to its consistency with the thermodynamic principle of superfluid flow.
The explicit form for the condensed boson velocity field means that
the chemical potential has gradient
\begin{equation}
\nabla \mu
=
- m  {\bf v}_0 \cdot  \nabla  {\bf v}_0
= -m \omega^2 [ x \hat{\bf x} + y\hat{\bf y}],
\end{equation}
or $\mu({\bf r}) = \mu({\bf 0}) - m\omega^2 r^2/2$.
(The vertical pressure gradient cancels with that of
the gravitational potential.)
This lateral gradient represents a centripetal force
that is directed toward the central axis.
(Number moves \emph{up} a chemical potential gradient.)
This centripetal force has the same magnitude and direction
as the force exerted on the normal fluid by the lateral pressure gradient.
The difference is that it is the chemical potential
that provides the driving force for the superfluid.
The thermodynamic principle of superfluid flow
is that it minimizes the energy at constant entropy (Attard 2025a \S4.6).
This means that it is the gradient of the chemical potential
that is the statistical force experienced by condensed bosons.
Hence the present result for the centripetal force
is entirely consistent with this principle.

\comment{ 
In the laboratory frame,
this force alters the trajectories of the condensed bosons
from straight lines to circular orbits.
This reduces the kinetic energy of a condensed boson
from what it would have been if it had continued straight
and acquired the velocity at a larger radius.
Since the chemical potential is the number derivative of the energy
at constant entropy,
the solution of rotating condensed bosons
maximizes the entropy of the universe.
This is demanded by the thermodynamic principle of superfluid flow
(Attard 2025a \S4.6).
This principle is deduced
from the H. London (1939) fountain pressure equation,
which itself has been confirmed by direct measurement.
The result is also demanded by the statistical mechanical theorem
that condensed bosons respond to applied forces
in inverse proportion to the occupancy of their momentum state,
which is Newton's second law of motion for an open quantum system,
Eq.~(\ref{Eq:NIIqu}) (Attard 2025b).
} 

\comment{ 
An interesting conceptual point is that since the superfluid
rotates in unison with the normal fluid it cannot be completely immune
to intermolecular forces.
This is consistent with Eq.~(\ref{Eq:NIIqu}),
namely that the response of  a single condensed boson to a force
is reduced, not eliminated.
Unless momentum transitions occur for sets of bosons,
it remains to reconcile this with the idea that superfluid flow has zero,
not merely reduced, viscosity.
} 

Arguably,
Landau's (1941) principle of irrotational superfluid flow
has become accepted simply by analogy
with classical inviscid flow
and with, in magnetic field free regions,
irrotational supercurrent flow
(but see \S\S\ref{Sec:Landau}, \ref{Sec:QuantCirc}, and \ref{Sec:MacroWave}).
In contrast,
the present conclusion that superfluid flow can be rotational
is based upon the facts that
(1) it correctly predicts the measured free surface
of a steadily rotating mixture of normal fluid and superfluid,
(2) it minimizes the energy at constant entropy,
which is the thermodynamic principle of superfluid flow
that is based upon the measured fountain pressure data (Attard 2025a \S4.6),
and (3) it conserves the molecular  probability distribution
for condensed bosons,
which is a requirement of equilibrium  statistical mechanics
(Attard 2025a Ch.~5, 2025b).

It was mentioned above that for superfluid and normal fluid mixtures
there is a qualitative difference
between transient or oscillating flow on the one hand,
and steady flow on the other.
Indeed there is ample experimental evidence
that the superfluid does not partake (or not fully partake)
in transitory or oscillatory rotational motion
(eg.\ Andronikashvilli 1946,
and the early measurements of the viscosity of helium~II).
The theoretical basis for this difference
lies in Newton's second law of motion for an open quantum system,
Eq.~(\ref{Eq:NIIqu})
(Attard 2025b),
which shows that applied forces do little to change the momenta
of condensed bosons.
In other words,
in the steady state the quantum second law has no direct effect
because the momentum is constant,
whereas in a transient or oscillatory system the applied forces
affect differently the condensed and the uncondensed bosons.
The present two-fluid equations
\emph{without} Landau's irrotational condition
could of course  be applied
to transient or to oscillating flows of helium~II .

\subsection{Irrotational Vortices and Rigid Rotation} \label{Sec:nonVortex}

Attempts to reconcile the predicted irrotational flow
with the measured rotational flow
are, in the opinion of the present author, unconvincing.
For example,
Pathria (1972 \S 10.7)
analyses a model that consists of a uniform distribution
of superfluid vortices,
which he attributes to Lane (1962)
(see also Landau and Lifshitz 1955, Lifshitz and Kagenov 1955,
Walmsley and Lane 1958).
The vortices have a particular form that is said to give them zero vorticity,
which is said to reconcile the predicted irrotational superfluid flow
with the measured rigid rotation discussed in the preceding subsection
(Annett 2004 \S 2.5, Pathria 1972  \S 10.7).

The so-called vorticity-free vortex,
in cylindrical coordinates ${\bf r} = \{\rho,\theta,z\}$,
is
\begin{equation}
{\bf v}({\bf r}) =
\left\{
\begin{array}{ll}
\displaystyle
\frac{K}{2\pi\rho}   \hat{\bm \theta}, & \rho > \rho_0 ,\\
\displaystyle \rule{0cm}{0.6cm}
\frac{K}{2\pi\rho_0} \hat{\bm \theta}, & \rho \le \rho_0 .
\end{array} \right.
\end{equation}
It is necessary to introduce the small radius cut-off $\rho_0$
to prevent the velocity diverging to infinity on the axis.
The curl of this is
\begin{equation}
\nabla \times {\bf v}({\bf r})
=
\frac{1}{\rho} \frac{\partial (\rho v_\theta)}{\partial \rho} \hat{\bf z}
=
\left\{
\begin{array}{ll}
\displaystyle
{\bf 0}, & \rho > \rho_0 ,\\
\displaystyle \rule{0cm}{0.6cm}
\frac{1}{\rho} \frac{K}{2\pi\rho_0} \hat{\bf z}, & \rho \le \rho_0 .
\end{array} \right.
\end{equation}
In the region where this vanishes, which is most of the range,
the flow is indeed irrotational.
But in the core region the curl is non-zero,
indeed divergent,
and the flow is definitely rotational.
To the present author it seems
a misnomer to call this a vortex with zero vorticity.

The circulation around a disc of radius $S \ge \rho_0$ enclosing the axis
of the so-called irrotational vortex is
\begin{eqnarray}
\oint {\rm d}{\bf l} \cdot {\bf v}({\bf r})
& = &
\int_{\rm S} {\rm d} {\bf S}  \cdot [\nabla \times {\bf v}({\bf r})]
\nonumber \\ & = &
2\pi \int_0^{\rho_0} {\rm d} \rho\, \rho
\frac{1}{\rho} \frac{K}{2\pi\rho_0}
\nonumber \\ & = &
K.
\end{eqnarray}
This is independent of the radius of the disc provided
that it is bigger than the cut-off
and that it encloses the axis of the vortex.

Pathria (1972 \S 10.7) invokes a uniform distribution
of this type of vortex, using the property of fixed circulation,
to conclude that the overall velocity field of the superfluid
is that of rigid rotation,
${\bf v}_0({\bf r}) = \omega \rho \hat{\bm \theta}$,
where $\rho$ is here measured from the axis
of the rotating system
(Annett 2004 Eq.~(2.46), Pathria 1972 Eqs~(10.7.7) and (10.7.9)).
This is certainly what is measured experimentally.
But there are two issues with the analysis.
First, it avoids any discussion of the core region of the vortices,
where the flow is certainly rotational,
which contradicts the assertion that superfluid flow must be irrotational.
And second, the final superfluid velocity field yielded by the model
is one of rigid rotation.
Surely this is in fact a simple and direct proof
that superfluid flow can be rotational.

\subsection{Landau's Irritational Principle for Superfluid Flow}
\label{Sec:Landau}

Landau (1941) developed
a principle of uniform irrotational superfluid flow
from the fundamental belief
that the superfluid state is isomorphic
with the ground energy  state.
Based on a quantum formulation of hydrodynamics (see \S\ref{Sec:QuantCirc}),
he purports to prove that in general
a uniform irrotational flow has lowest energy,
with an energy gap to rotational flow.
He concludes:
`The supposition that the normal level of potential motions
[ie.\ $\nabla \times {\bf v}_0 = {\bf 0}$]
lies lower [in energy]
than the beginning of the spectrum of vortex motions
[ie.\ $\nabla \times {\bf v}_0 \ne {\bf 0}$]
leads to the phenomenon of superfluidity' (Landau 1941 p.~356).

However, even if uniform irrotational flow
has the lowest energy with a gap,
the more numerous states with rotational flow
are likely occupied for entropic reasons,
particularly given that the measurements are so far from absolute zero.
Further, since a macroscopic amount of $^4$He is involved in superfluid flow,
at these temperatures on the order of 50\% of the total number,
these cannot all be in the ground energy state.
Landau's (1941) assertion
that superfluid flow must be irrotational everywhere
is based on the unsound supposition that superfluid helium
must be in the ground energy state.

In addition, for the non-equilibrium case of forced flow,
such as the driven rotation of a bucket of saturated liquid helium
below the $\lambda$-transition,
analysis based on the equilibrium occupation of energy states
is largely irrelevant.
Landau's (1941 p.~357) conclusion
`when the walls of the vessel are in motion,
only a part of the mass of liquid helium is carried along by them,
and the other part remains stationary' has not aged well.

More generally, the present author deprecates Landau's (1941) project
to formulate a theory of quantum hydrodynamics
based on `macroscopic wave functions'
that obey quantum mechanics (see \S\ref{Sec:MacroWave}).
The present author judges this to be fundamentally incompatible
with the principles of non-equilibrium thermodynamics (Attard 2012)
and with the principles of quantum statistical mechanics (Attard 2021).

Finally, Landau (1941) rejected F. London's (1938) theory
that the $\lambda$-transition and superfluidity
in $^4$He was a consequence of Bose-Einstein condensation:
`Tisza's' [ie.\ F. London's (1938)] well-known attempt to consider
helium II as a degenerate Bose gas cannot
be accepted as satisfactory' (Landau 1941 p.~356).
Amusingly, even today researchers continue to embrace both theories
despite their  mutual contradiction.
Bogulbov (1947) attempted to reconcile the two theories;
the common element in all three theories is the assumption
that superfluidity was due to $^4$He being in the energy ground state.

It was mentioned above that even if irrotational flow
was the lowest energy flow,
this would not imply that superfluid flow had to be irrotational.
This logical inconsistency underlying Landau's (1941) principle
appears to have been recognized by Feynman (1954 p.~276),
 who says
`The third [question] is to describe states for which the superfluid velocity
is not vortex free. 
\ldots
A new element must presumably be added to our picture
[of vortex free motion]'.
Thus Feynman appreciates that
irrotational motion describes only some superfluid flow,
and that rotational flows are allowed.

\subsection{Quantized Circulation} \label{Sec:QuantCirc}

Hydrodynamics is defined over macroscopic volumes
that are large on the molecular scale
but small on the scale of variations in the fluxes and thermodynamic fields.
The superfluid velocity ${\bf v}_0({\bf r})$ is such a hydrodynamic flux,
which is to say that it is the result of averaging the velocities
of a macroscopic number of $^4$He atoms
in a volume about ${\bf r}$.

Onsager (1949)
and Feynman (1955)
argued that
the superfluid underwent quantized vortex motion.
Accordingly,
for a ring of bosons of macroscopic size,
the quantized circulation in a superfluid is
(Pathria 1972 Eq.~(10.6.7))
\begin{equation}
\oint {\rm d}{\bf l} \cdot {\bf v}_0({\bf r})
=
\int_{\rm S} {\rm d} {\bf S}  \cdot (\nabla \times {\bf v}_0({\bf r}))
= \frac{2\pi \hbar n}{m} ,
\end{equation}
where $n$ is meant to be an integer.
This says that the curl of the velocity field
through the area of integration must be quantized.
Pathria (1972 \S10.6) argues that
if the area of integration is shrunk continuously,
then the right hand side would change discontinuously unless $n=0$.
Therefore it must be the case that
$\nabla \times {\bf v}_0({\bf r}) = {\bf 0}$
for all radii.
Thus quantization is said to prove that superfluid flow must be irrotational.

However, the issue is that since $ {\bf v}_0({\bf r})$
is the macroscopically averaged hydrodynamic velocity field,
the quantum number $n$ must also be an average,
which means that it belongs to the continuum.
(For example, the average number of bosons in a given volume
is not an integer.)
Hence the right hand side can change continuously
as the area of integration is changed.
Hence there is no reason to insist that $n=0$,
and there is no proof on the basis of quantization that
$\nabla \times {\bf v}_0({\bf r}) = {\bf 0}$.

\subsection{Macroscopic Wavefunction} \label{Sec:MacroWave}

Related to  Landau's (1941)
formulation of quantum hydrodynamics
is the idea of a macroscopic wavefunction
defined in three-dimensional position space.
This is said to obey the operator relationships
of a normal quantum wavefunction (eg.\ Annett 2004 \S2.4).
Identifying the modulus squared with the condensed boson density
(Annett 2004 \S2.3, Pathria 1972, \S10.5),
gives superfluid flow
proportional to the gradient of a velocity potential,
which means that it is irrotational.

There are several objections to the macroscopic wavefunction approach.

First, according to Pathria (1972 \S 10.5),
it is a sort of ideal gas approximation
in which the energy eigenfunction of the whole system
is factorized as the product of single-particle energy eigenfunctions.
Although this is argued as a type of mean field approximation,
it is a dubious approach to condensed matter
as it  ignores molecular structure and correlations between the particles
due to their interactions
(see point three).

Second,
it applies only to the ground state.
This restriction is necessary because the factorized product
has to consist of identical energy eigenfunctions,
namely the macroscopic wavefunction,
and the ground state is the only state in which the particles
are in the same single-particle state
(Pathria 1972 \S 10.5).
The present author has presented extensive evidence
that Bose-Einstein condensation is not solely into the ground energy state
(Attard 2025a \S\S1.1.4 and 2.5).
In short,
it strains credulity to assert that the superfluid
and superconductor transitions,
which occur at temperatures far above absolute zero,
are dominated solely by particles in the energy ground state.

Third,
identifying the square of the amplitude of the macroscopic wave function,
$|\psi_0({\bf r})|^2$,
with the condensed boson density, $n_0({\bf r})$, is unrealistic.
It is not immediately obvious what
the single-particle mean-field ground-state energy eigenfunction
has to do with density.
Perhaps one might argue that,
apart from normalization,
both the Born probability,  $|\psi_0({\bf r})|^2$,
and the single particle density, $n_0({\bf r})$,
give the probability of finding a particle
at ${\bf r}$ irrespective of the other particles.
But the macroscopic wave function formulation
implies that the $m$-particle density is
$n^{(m)}_0({\bf r}_1,{\bf r}_2,\ldots,{\bf r}_m)
= \prod_{j=1}^m n_0({\bf r}_j)$,
which is a very poor approximation for a condensed liquid
because it neglects correlations and attractions between the particles,
and it allows particles to overlap.

Fourth,
the evolution of the macroscopic wave function
is governed by the Schr\"odinger equation,
specifically the non-linear mean-field version
due to Ginzburg and Pitaevskii (1958)
and to Gross (1958, 1960).
This implies that quantum mechanics also governs its flux,
as in
(Annett 2004 Eq.~(2.21), Tinkham 2004 Eq.~(4.14))
\begin{equation} \label{Eq:J0}
{\bf J}_0({\bf r})
 =  \frac{-{\rm i}\hbar}{2m}
[ \psi_0({\bf r})^*\, \nabla \psi_0({\bf r})
- \psi_0({\bf r}) \nabla \psi_0({\bf r})^*\,].
\end{equation}
This approach,
with the condensed boson density replacing the macroscopic wave function,
is sometimes praised as the ground breaking basis for quantum hydrodynamics.
But quantum mechanics is fundamentally incompatible with hydrodynamics:
the former applies to a few individual particles isolated
from their surroundings,
whereas the latter deals with a macroscopic numbers of particles
contained in local volumes that are large on the molecular scale
but small on the scale of variations in the fields such as velocity,
temperature, pressure, etc.
The hydrodynamic flux on the left hand side of this equation
has nothing to do with the single-particle quantum dynamics
on the right hand side.
Rather it should be the average of the right hand side
over a macroscopic number of non-identical, interacting wavefunctions.
Indeed, quantum statistical mechanics
shows that in the case of large numbers of interacting particles,
the Schr\"odinger equation has to be replaced
by an equation that takes into account the wavefunction collapse
due to interactions with the environment
(Attard 2025b).

And fifth,
the macroscopic wave function
has been given several, contradictory, microscopic interpretations:
is it the gap parameter in BCS theory
(Gor'kov 1959, Tinkham 2004 \S1.5)?
Or is it the condensed boson density
(Annett 2004 \S2.3, Pathria 1972 \S10.5, Tinkham 2004 \S1.5)?
Or is it the single-particle mean-field ground-state energy eigenfunction
(Pathria 1972 \S10.5)?
The plethora of differing explanations
suggests that in fact it has no convincing basis in reality.

To be clear, the present author
does not object to using the condensed boson density
$n_0({\bf r})$ as the order parameter
in Landau's (1937) phenomenological theory of second order phase transitions.
This is the density of condensed $^4$He for superfluidity,
and the density of condensed bosonic electron pairs
for superconductivity.
The objection is to the macroscopic wave function $\psi_0({\bf r})$
and to its use in the quantum flux equation
to predict superfluid or superconductor flow.
Of course if $n_0({\bf r})$ is used directly as the order parameter,
then there is no recourse to the  Schr\"odinger equation
or the quantum flux equation.
In this case the order parameter approach \emph{per se} provides no basis
for the dynamics of superfluid flow or for supercurrents.

There are several predictions of the macroscopic wave function
that directly contradict measured data.

First,
using the macroscopic wavefunction,
$\psi_0({\bf r}) = \sqrt{ n_0({\bf r})} \, e^{{\rm i}\theta({\bf r})}$,
in the quantum mechanical   expression for the flux,
Eq.~(\ref{Eq:J0}),
gives
(Annett 2004 Eq.~(2.21))
\begin{eqnarray}
{\bf J}_0({\bf r})
& = &
 \frac{\hbar}{m} n_0({\bf r}) \nabla \theta({\bf r}).
\end{eqnarray}
This says that the local superfluid velocity is  proportional
to the gradient of the phase of the macroscopic wave function,
${\bf v}_0({\bf r}) = (\hbar/m) \nabla \theta({\bf r})$.
If true, as the gradient of a scalar
this would make superfluid flow irrotational,
$\nabla \times {\bf v}_0({\bf r}) = {\bf 0}$.
But, as discussed in detail above,
measurement shows
that superfluid flow has non-zero rotation
(Osborne 1950, Walmsley and Lane 1958).

Second,
in the case of superconductivity
it predicts a temperature scaling
$n_{20}(T) \sim  [1-T/T_{\rm c}]$ (Tinkham 2004 Eq.~(4.6)).
But the experimentally measured temperature dependence
of the London penetration length is
$ \lambda(T) \sim \lambda(0) [1-(T/T_{\rm c})^4]^{-1/2}$
(Tinkham 2004 Eq.~(1.7)),
which implies that $n_{20}(T) \sim [1-(T/T_{\rm c})^4]$.
There is a clear contradiction between the predicted and the measured
temperature dependence
that is only resolved in the limit $T \to T_{\rm c}^-$.
This is a greatly restrictive limit
that casts doubt on the physical reality of equating $|\psi_0({\bf r})|^2$
with $n_{20}({\bf r})$ in the condensed bosonic electron pair regime.
It means that the condensed boson velocity field
predicted by the quantum flux equation
does not apply anywhere in the condensed regime
except possibly in the immediate vicinity of the transition,
$T \to T_{\rm c}^-$.

This second point for superconductivity is  a reflection
on the limits of Landau's (1937) phenomenological theory.
Actually for superconductivity
it is necessary that $\nabla \theta({\bf r}) = {\bf 0}$
for consistency with the second London equation and the London gauge,
in which case the macroscopic wave function reduces
to the square root of the condensed bosonic pair density.

\subsection{Path Integral Monte Carlo}

Path integral Monte Carlo is a method for simulating
quantum systems (Allen and Tildesley 1987).
Ceperley and co-workers have introduced various algorithms,
and their application to  liquid helium has been reviewed
(Ceperley 1995).
Here the method is critically assessed,
the excuse being the moment of inertia of $^4$He,
which  below the $\lambda$-transition is found to be less than that expected
for the entire fluid (Ceperley 1995 \S IIIE).
The interpretation of this is that the superfluid component
does not participate in rotation,
and that the fractional decrease in the moment of inertia
may be used to give an estimate of the superfluid fraction
that complements other methods of estimation.
The quantitative values for the superfluid fraction will be addressed below,
while  here it is emphasized that there is no inconsistency
between this reduction in the moment of inertia
and the above conclusion that superfluid flow need not be irrotational.
The moment of inertia is relevant for transient or oscillatory
rather than steady rotation,
and it is known experimentally
that the superfluid does not respond to time-varying forces
(Andronikashvilli 1946)
whilst it fully participates in steady rotation (Osborne 1950).

Early on Feynman (1953)
mapped imaginary time path integrals of a quantum system onto
a classical system of interacting ring polymers
He showed that the Bose condensation occurred at  temperatures lower
for interacting helium than for free particles.
Ceperley (1995 p.~282) observes that
`path integrals do not naturally describe
the quasiparticle picture of liquid helium, as a gas
of interacting phonons and rotons [ie.\ Landau's (1941) theory].
Feynman switched
away from the path-integral theory of helium in favor
of the excitation picture: phonons, rotons, and vortices'.

The path integral Monte Carlo method
defines a high temperature $\tau \equiv \beta/M$
(the imaginary time step),
and writes the density matrix as
(Ceperley 1995 Eq.~(2.9))
\begin{eqnarray}
\lefteqn{
\rho(R_0,R_M;\beta)
} \nonumber \\
& = &
\int {\rm d}R_1\,{\rm d}R_2 \ldots{\rm d}R_{M-1}\;
\rho(R_0,R_1;\tau)
\nonumber \\ && \mbox{ } \times
\rho(R_1,R_2;\tau)\ldots\rho(R_{M-1},R_M;\tau) .
\end{eqnarray}
Here $R_m = \{ {\bf r}_{1,m},{\bf r}_{2,m},\ldots,{\bf r}_{N,m}\}$
is the position configuration at temperature-slice $m$.
Evaluating  these configurations at the higher temperature $\tau$
may be viewed as a form of umbrella sampling.

With the Hamiltonian operator being the sum of the kinetic and potential
energy operators,
$\hat{\cal H} = \hat{\cal T} + \hat{\cal V}$,
 the commutator in the operator identity,
$
e^{-\tau(\hat{\cal T} + \hat{\cal V})
+ \tau^2[\hat{\cal T} , \hat{\cal V}]/2}
=
e^{-\tau \hat{\cal T}} e^{-\tau \hat{\cal V} }$,
can be neglected for large $M$. 
The primitive approximation is
\begin{eqnarray}
\rho(R_0,R_2;\tau)
& \approx &
\int {\rm d}R_1 \;
\langle R_0 | e^{-\tau \hat{\cal T} } | R_1 \rangle
\langle R_1 | e^{-\tau \hat{\cal V} } | R_2 \rangle
\nonumber \\ & = &
\frac{e^{-(R_0-R_2)^2/4\lambda\tau}}{(4\pi\lambda\tau)^{3N/2}}
\, e^{\beta V(R_2)}.
\end{eqnarray}
Here the  sum over momenta
has been approximated by a continuum integral,
which gives the Gaussian links between the ring polymer beads,
$4\lambda\tau = 4 \hbar^2 \beta/2m M =\Lambda^2 /M\pi$,
where $\Lambda$ is the usual thermal wavelength.
With this the density matrix is
\begin{eqnarray}
\rho(R_0,R_M;\beta)
& = &
\int {\rm d}R_1\ldots{\rm d}R_{M-1} \;
 \\ &  & \mbox{ } \times \nonumber
\prod_{m=1}^M
\frac{ e^{-(R_{m-1}-R_m)^2/4\lambda\tau} }{ (4\pi\lambda\tau)^{3N/2} }
e^{-\tau V(R_m) } .
\end{eqnarray}
The diagonal element of the density matrix
describes a classical system of ring polymers,
with harmonic intrapolymer interactions 
between adjacent slices,
and the usual interactions between atoms within each slice.
The quantization of momenta has been neglected.

Ceperley (1995 Eq.~(2.28)) accounts for wave function symmetrization
by defining the bosonic density matrix,
\begin{equation}
\rho_{\rm B}(R_0,R_1;\beta)
=
\frac{1}{N!} \sum_{\hat{\rm P}}
\rho(R_0,\hat{\rm P}R_1;\beta) ,
\end{equation}
where the permutator of $N$ ordered objects appears.
Successful transpositions cross-link pairs of polymers creating
larger ring polymers.
In the practical implementation of the path integral method,
Ceperley (1995) explores cross-linking between up to four ring polymers,
for typical system sizes of $N=64$ and $\tau=40$\,K$^{-1}$
(and $M=$ 20--1000, depending on the particular algorithm).

One may question whether the random walk through permutation space
is particularly efficient or ergodic,
and whether  a maximum cycle (loop) length of 4 is sufficient.
For comparison, the classical phase space simulations
of Attard (2025a, 2025d) used $N=5,000$ and loop lengths up to seven
in an exponential resummation (ie.\ more like $e^7$).
For each state point the path integral algorithm
took about an hour on a 1986 Cray-I supercomputer (Ceperley 1995),
whereas the classical phase space algorithm
took about 48 hours on a 2007 Windows personal computer
(Attard 2025a, 2025d).

The heat capacity in the neighborhood of the $\lambda$-transition peak
is reproduced accurately on the low temperature side
(Ceperley 1995 Fig.~11, Ceperley and Pollock 1986 Fig.~1).
On the high temperature side an outlier point suggests
systematic errors of unknown origin
that are larger than the estimated statistical error.
Possibly these are related to the limitations of the permutation algorithm
discussed above.
Attard (2025d) found that the heat capacity diverged on the high temperature
side of the $\lambda$-transition due to the growth in size and number
of position permutation loops.
It is difficult to see how this divergence could be captured
in the path integral simulations when the maximum loop size
is restricted to $4$, and only $N=64$ atoms are used.

It should be pointed out that the path integral method locates
the $\lambda$-transition almost exactly
at the experimental value of $T_\lambda=2.17$\,K
(Ceperley 1995 Fig.~11), 
whereas the classical phase space method locates it at
$T_\lambda \alt 3.5$\,K (Attard 2025d).
The better performance of the path integral method
by this measure  likely lies in its
essentially exact treatment of the commutation relation
for the position and momentum operators.
The classical phase space method in the current implementation
(Attard 2025a, 2025d)
neglects the commutation function on the grounds
that it is short-ranged.

Ceperley (1995) gives several methods to estimate
the condensate (equivalently, superfluid) fraction.
These are (1) a winding number method,
(2) a method based on the number of atoms with zero momentum,
and (3) a method based on the moment of inertia.

The winding number method counts 
ring polymers that span the system.
This definition of condensation appears to be motivated
by Feynman's (1953) observation that
`the superfluid transition
is represented\ldots 
by the formation of macroscopic polymers, i.e., those stretching across
an entire system' 
(Ceperley 1995 p.~290).
Ceperley (1995 p.~340) describes the winding number method
as `the most elegant approach',
but it seems to the present author that it is the least justified
and the worst performing of the three methods.
Above the $\lambda$-transition temperature,
at $T=2.22$\,K it gives the superfluid fraction as $N_{\rm s}/N = 0.32(4)$
(Ceperley 1995 Fig.~49), which is absurdly high.
This was the result of about $3\times 10^4$ successful permutation moves
(spread amongst $5\times 10^6$ attempted path moves, 
taking 50~hrs on a workstation),
which appears too small to be a representative sample of the
$64! = {\cal O}(10^{100})$ possible permutations of each configuration.
An obvious problem with this approach
is that it can be expected to depend upon the simulation parameters:
the proportion of ring polymers that span the system
decreases with increasing system size,
and increases with increasing maximum cycle (loop) size.

In the opinion of the present author,
condensation and permutation are not equivalent.
For example, the present author finds a divergence
of the heat capacity on the high temperature side
of the $\lambda$-transition
due to the divergence in size and number of position permutation loops
(Attard 2025a).
In this regime the present author finds zero condensation (Attard 2025d),
and the experimentalists find zero superfluidity
(Donnelly and  Barenghi  1998).

The second method for estimating condensation 
defines `the condensate fraction as the probability of finding
an atom with precisely zero momentum' (Ceperley 1995 p.~297).
This appears motivated by the usual interpretation
that Bose-Einstein condensation is into the energy ground state
(Einstein 1924, 1925, F. London 1938),
an interpretation that the present author deprecates (Attard 2025a).
(Onsager and Penrose (1956) argue for condensation into a single
non-zero single-particle state.)
In the path integral method
either the asymptotic behavior, or else the zero moment,
of the off-diagonal single particle density matrix are used,
since these are related to its Fourier transform at zero momentum.
These methods give a non-zero condensate fraction
above the $\lambda$-transition temperature (Ceperley  and Pollock  1986),
which contradicts the experimental evidence that there is none.
Ceperley (1995) attributes this to the finite size of the simulated system.
The present author believes that more likely
 this definition of condensation is unphysical.

A more serious example of unphysical behavior with this second method
is that below the $\lambda$-transition
the maximum fraction of condensed bosons
in the path integral simulations
was always less than 10\%.
This is consistent with estimates by others:
Penrose and Onsager (1956) using Feynman's approximation,
McMillan (1965) using Monte Carlo calculations,
and Kalos \emph{et al.}\ (1981) and Whitlock and Panoff (1987)
using Green's-function Monte Carlo calculations
all estimate the ground-state condensate as close to 8--9\%
(Ceperley 1995 p.~18).
This is again in obvious contradiction to the evidence:
experimentally the fraction of superfluid (equivalently, condensed) atoms
goes to unity as $T\to0$ (Donnelly and  Barenghi  1998),
and thermodynamically the fraction of atoms in the ground momentum state
goes to zero as $N\to\infty$ (Attard 2025a \S\S 1.1.4 and 2.5).

Ceperley (1995 p.~297) says
`One often hears the question: why is the condensate
fraction at low temperature only 10\% while the system
is 100\% superfluid?'.
Unfortunately, in answer he only explains
the first part as a consequence of the particular path integral algorithm
without relating this to the second part of the question.
Other attempts to explain away this contradiction
are somewhat convoluted (eg.\ Annett 2004 \S2.7).
The present author believes
that the obvious answer is the correct one:
condensation is not solely into the ground state.

The third path integral Monte Carlo method
to estimate the condensation (superfluid) fraction
is from the moment of inertia.
The fractional reduction in this from that expected for the entire liquid
is taken to equal the superfluid fraction.
This method yields realistic results at and below the $\lambda$-transition,
approaching unity for $T \ll T_\lambda$,
but it also yields a non-zero fraction of condensed bosons
above the $\lambda$-transition
(Ceperley 1995 Fig.~26).
The amount of condensation (superfluid) is as high as 20\% in this regime,
whereas the measured value is zero  (Donnelly and  Barenghi  1998).
It appears that this problem arises from using the continuum representation
of momentum in the path integral approach,
since this precludes the competition
between position and momentum permutation loops
that occurs for quantized momentum
and that suppresses condensation on the high temperature side
of the $\lambda$-transition
(Attard 2025d).

The present author defines condensation 
as the multiple occupancy of multiple low-lying momentum states.
He believes that the problems with condensation
in the path integral method stem from the definitions
---condensation is any permuted ring polymer that winds around the system,
or else condensed bosons must have zero momentum---
and also from the invocation of  the momentum continuum.
It is these that are irresponsible for the unphysical results
 for condensation given by the path integral method(Ceperley  1995).


\subsubsection{Other Simulation Algorithms}

Apart from the author's own classical phase space quantum Monte Carlo
(CPSQMC)
(Attard 2018b, 2025a Ch.~3, 2025d)
and classical phase space quantum molecular dynamics
(CPSQMD) (Attard 2025b) methods,
the main computer simulation methods for quantum systems
that compete with path integral Monte Carlo (PIMC)
are variational methods.
The following summary is taken from Ceperley (1995)
and the reader is referred to that source for greater detail.

Variational Monte Carlo (VMC),
which minimizes the energy based on a trial wave function,
has been applied to helium
(McMillan 1965, Ceperley and Kalos 1979, Schmidt and Ceperley 1992).
Although an order of magnitude faster than path integral Monte Carlo,
the quality of the results depends upon
the skill in selecting a trial wave function,
and it is applicable only at zero temperature.
Variational path integrals (VPI)
(Reatto and Masserini 1988, Vitiello, Runge, and Kalos, 1988),
and shadow wave functions (SWF) (Vitiello \emph{et al.}\ 1988, 1990)
have some advantages over VMC,
although there remain problems with simulating excited states.
These have also been applied to helium.

The Green's function Monte Carlo (GFMC) method
was first applied to liquid helium by
Kalos, Levesque, and Verlet (1974).
Reviews of this method and the closely related
diffusion Monte Carlo method
have been given by  Ceperley and Kalos (1979) and
Schmidt and Ceperley (1992).
At each step in the process,
an ensemble of several hundred configurations is evolved
by diffusion, duplication, and deletion.
The method has problems with large systems
(the time step goes as $N^{-1/2}$),
with highly correlated ensemble members,
and with wave function symmetrization,
but has better ergodicity than PIMC.

Ceperley (1995 p.~348) concludes that
`For all of these reasons, PIMC is a better ``black box"
[ie.\ less subjective] than VMC, VPI, or GFMC.'

Finally, effective-potential Monte Carlo
uses a quadratic trial potential
to minimize the free energy (Liu \emph{et al.}\ 1991, 1992).
It appears most useful for nearly classical systems
that are nearly harmonic.

%
\section{Conclusion}
\setcounter{equation}{0} \setcounter{subsubsection}{0}
\renewcommand{\theequation}{\arabic{section}.\arabic{equation}}
%

This paper has shown that the two-fluid theory
for superfluid hydrodynamics
derives from the fountain pressure equations,
(\ref{Eq:HLondon}) and  (\ref{Eq:muA=muB}),
namely that condensed bosons move at constant entropy.
But what is the molecular basis for these equations?

For an open quantum system,
the average rate of change of momentum
for boson $j$ in the momentum state ${\bf a}$
experiencing the classical force ${\bf f}_{j}$ is  (Attard 2025b)
\begin{equation}  \label{Eq:NIIqu}
\left\langle \dot {\bf p}_j \right\rangle
=
\frac{1}{ N_{\bf a} } {\bf f}_{j} .
\end{equation}
This differs from Newton's second law of motion
by the occupancy,
$N_{\bf a} = \sum_{j=1}^{N} \delta_{{\bf p}_j,{\bf a}}$.
In the classical regime there are many more accessible momentum states
than that are bosons,
so that boson $j$ is the sole occupant of its momentum state,
$N_{\bf a}=1$.
Conversely,
in the quantum regime condensed bosons in highly occupied momentum states,
$N_{\bf a}  = {\cal O}(10^3)$,
are insensitive to molecular forces
and  they rarely change their momentum state.
This explains the reduction in viscosity in superfluid flow.
This result preserves the equilibrium probability distribution,
which is the exponential of the entropy,
including the occupation entropy due to Bose-Einstein condensation,
$S^{\rm occ} = k_\mathrm{B} \sum_{\bf a} \ln  N_{\bf a}!$ (Attard 2025b).

A serious misunderstanding in superfluidity is the assertion that
condensed bosons have zero entropy (H. London 1939, Tisza 1947).
In fact the occupation entropy of condensed bosons is
so much greater than the configurational entropy
that it restricts their motion to lines of constant entropy.
This is the real origin
of H. London's (1939) fountain pressure equation 
and of the principle of superfluid flow,
namely that the motion of condensed bosons preserves the entropy
(Attard 2025b).

A second conceptual error is
invoking a macroscopic wavefunction
as an order parameter, as the basis of quantum hydrodynamics,
and as the proof that superfluid flow is irrotational (Landau 1941).
This lead to the idea that the induction of vortices with non-zero vorticity
destroys superfluidity (Feynman 1955).
In fact superfluid flow can have non-zero vorticity,
as was was explicitly demonstrated in \S\ref{Sec:Revolve}
for steady rotation.

A third error is Einstein's (1924, 1925),
that condensation is solely into the ground state.
The fruit of that poisonous tree contaminates the field
to this day.

\section*{References}


\begin{list}{}{\itemindent=-0.5cm \parsep=.5mm \itemsep=.5mm}

\item 
Allen M P and Tildesley D J 1987
\emph{Computer Simulation of Liquids}
(Oxford: Clarendon Press)

\item 
Andronikashvilli E L 1946
\emph{J.\ Phys.\ USSR} {\bf 10} 201

\item 
Annett J E 2004
\emph{Superconductivity, Superfluids and Condensates}
(Oxford: Oxford University Press)

\item %
Attard P 2002
\emph{Thermodynamics and Statistical Mechanics:
Equilibrium by Entropy Maximisation}
(London: Academic)

\item 
Attard  P 2012
\emph{Non-equilibrium Thermodynamics and Statistical Mechanics:
Foundations and Applications}
(Oxford: Oxford University Press)

\item 
Attard P 2018b
quantum statistical mechanics in classical phase space. Expressions for
the multi-particle density, the average energy, and the virial pressure.
arXiv:1811.00730

\item 
Attard P  2021
\emph{Quantum Statistical Mechanics in Classical Phase Space}
(Bristol: IOP Publishing)

\item 
Attard P 2025a
\emph{Understanding Bose-Einstein Condensation,
Superfluidity, and High Temperature Superconductivity}
(London: CRC Press)

\item
Attard P 2025b
The molecular nature of superfluidity: Viscosity of helium from quantum
stochastic molecular dynamics simulations over real trajectories
arXiv:2409.19036v5

\item 
Attard P 2025d
Bose-Einstein Condensation and the Lambda Transition
for Interacting Lennard-Jones Helium-4
arXiv:2504.07147v1

\item 
Bogolubov N 1947
On the theory of superfluidity
\emph{J. Phys.}\ {\bf 11} 23

\item 
Ceperley  D M  1995
Path integrals in the theory of condensed helium
\emph{Rev.\ Mod.\ Phys.}\ {\bf 67} 279

\item 
Ceperley D M and Kalos M H 1979
Quantum Many-Body Problems
in \emph{Monte Carlo Methods in Statistical Physics},
edited by K. Binder (Heidelberg: Springer p.~145)

\item 
Ceperley D M and Pollock E L 1986
Path-integral computation of the low-temperature properties of liquid $^4$He
\emph{Phys.\ Rev.\ Lett.}\ {\bf 56} 351

\item 
Donnelly R J and  Barenghi C F 1998
The observed properties of liquid Helium at the saturated vapor pressure
\emph{J.\ Phys.\ Chem.\ Ref.\ Data} {\bf 27} 1217

\item 
Donnelly R J 2009
The two-fluid theory and second sound in liquid Helium
\emph{Physics Today} {\bf 62} 34

\item 
Einstein A 1924
Quantentheorie des einatomigen idealen gases
Sitzungsberichte der Preussischen Akademie der Wissenschaften
{\bf XXII} 261 

\item 
Einstein A  1925
Quantentheorie des einatomigen idealen Gases. Zweite abhandlung.
Sitzungsberichte der Preussischen Akademie der Wissenschaften
{\bf I} 3

\item 
Fetter A L 1963
Vortex rings and the critical velocity in helium II
\emph{Phys.\ Rev.\ Lett.}\ {\bf 10} 507

\item 
Feynman R P 1953
\emph{Phys.\ Rev.}\ {\bf 90} 1116,
{\bf 91} 1291,  1301.

\item 
Feynman R P 1954
Atomic theory of the two-fluid model of liquid helium
\emph{Phys.\ Rev.}\ {\bf 94} 262

\item 
Feynman R P 1955
\emph{Progress in Low Temperature Physics}
ed.\ C J Gorter
(Amsterdam: North Holland) {\bf 1} 17

\item 
Ginzburg V L and Pitaevskii L P 1958
\emph{Zh.\ Eksperim.\ i.\ Teor.\ Fiz.}\ {\bf 34} 1240.
\emph{Sov.\ Phys.\ JETP} {\bf 7} 858).

\item 
Gor'kov L P 1959
\emph{Zh.\ Eksperim.\ i.\ Teor.\ Fiz.}\ {\bf 36} 1918.
Microscopic Derivation of the Ginzburg-Landau Equations
in the Theory of Superconductivity
\emph{Sov.\ Phys.\ JETP} {\bf 9} 1364.

\item 
de Groot S R and Mazur P 1984
\emph{Non-equilibrium Thermodynamics}
(New York: Dover)

\item 
Gross E P 1958
Classical theory of boson wave field
\emph{Annals of Phys.}\ {\bf 4} 57

\item 
Gross E P 1960
Quantum theory of interacting bosons
\emph{Annals of Phys.}\ {\bf 9} 292

\item 
Hammel E F  and Keller W E 1961
Fountain pressure measurements in liquid He II
\emph{Phys.\ Rev.}\ {\bf 124} 1641

\item 
Kalos M H Lee M A  Whitlock P A and Chester G V 1981
Modern potentials and the properties of condensed He 4
\emph{Phys.\ Rev.}\ B {\bf 24} 115

\item 
Kalos M H Levesque D and Verlet L 1974,
Helium at zero temperature with hard-sphere and other forces
\emph{Phys.\ Rev.}\ A {\bf 9} 2178

\item 
Kawatra M P and Pathria R K 1966
Quantized vortices in an imperfect
Bose gas and the breakdown of superfluidity in liquid helium II
\emph{Phys.\ Rev.}\ {\bf 151}  132

\item 
Landau L D 1937
On the theory of phase transitions
\emph{Zh.\ Eksperim.\ i.\ Teor.\ Fiz.}\ {\bf 7} 19

\item 
Landau L D 1941
Theory of the superfluidity of helium II
\emph{Phys.\ Rev.}\ {\bf 60} 356.
Two-fluid model of liquid helium II
\emph{ J.\ Phys.\ USSR} {\bf 5} 71

\item 
Landau L D and  Lifshitz E 1955
\emph{Doklady Akad.\ Nauk USSR} {\bf 100} 669

\item 
Lane C T 1962
\emph{Superfluid Physics} (McGraw-Hill: New York)

\item 
Lifshitz E  and Kaganov M 1955
\emph{J.\ Exptl.\ Theoret.\ Phys.\ USSR} {\bf 29} 257.
Translation: \emph{Soviet Phys.\ JETP} {\bf 2} 172

\item 
Liu S Horton G K and Cowley E R 1991
Variational path-integral theory of thermal properties of solids
\emph{Phys.\ Rev.}\ B {\bf 44} 11714

\item 
Liu S Horton G K  Cowley E R
McGurn A R Maradudin A A and Wallis R F 1992
Comparative study of a model quantum solid using quantum Monte Carlo,
the effective potential, and improved self-consistent theories
\emph{Phys.\ Rev.}\ B {\bf 45} 9716

\item  
London F 1938
The $\lambda$-phenomenon of liquid helium and the Bose-Einstein degeneracy
\emph{Nature} {\bf 141} 643

\item 
London F and London H 1935
The electromagnetic equations of the supraconductor
\emph{Proc.\ Royal Soc.\ (London)} {\bf A149}, 72.
Supraleitung und diamagnetismus
\emph{Physica} {\bf 2}, 341. 

\item 
London H 1939
Thermodynamics of the thermomechanical effect of liquid He II
\emph{Proc.\ Roy.\ Soc.}\ {\bf  A171} 484

\item 
McMillan W L 1965
Ground State of Liquid $^4$He
\emph{Phys.\ Rev.}\ A {\bf 138} 442

\item  
Onsager L 1949
Statistical Hydrodynamics
\emph{Nuovo Cim.}\ {\bf 6} Suppl.~2 279

\item  
Osborne D V 1950
The rotation of liquid helium II
\emph{Proc.\ Phys.\ Soc.\ London} A {\bf 63} 909.

\item 
Pathria R K 1972
\emph{Statistical Mechanics} (Oxford: Pergamon Press)

\item  
Penrose O and Onsager L 1956
Bose-Einstein Condensation and Liquid Helium
\emph{Phys.\ Rev.}\ {\bf 104} 576.
This paper relies on the axiom that
`It is characteristic of an ideal B.E. gas in equilibrium
below its transition temperature that a finite [ie.\ non-zero] fraction
of the particles occupies the lowest single-particle
energy level' (p.~577).
But the known analytic result for ideal bosons
gives the ground state occupancy as intensive,
$\overline N_{\bf 0}=e^{\beta\mu}/(1-e^{\beta\mu})$
(Attard 2025a Eq.~(2.10), Pathria 1972 Eq.~(7.1.22)).
In the thermodynamic limit the fraction vanishes,
$\overline N_{\bf 0}/N \to 0$.
Similarly, intensivity holds for 
each momentum state,
and also for interacting particles.
Hence the conclusion of Penrose and Onsager,
`B.E. condensation \ldots is present
whenever a finite fraction\ldots of the particles
occupies one single-particle quantum state' (p.~583),
is, in one word, wrong.

\item 
Reatto L and Masserini G L 1988
Shadow wave function for many-boson systems
\emph{Phys.\ Rev.}\ B {\bf 38} 4516

\item 
Schmidt K E and Ceperley D M 1992,
in \emph{Monte Carlo Methods in Condensed Matter Physics}
edited by K. Binder \emph{Topics in Applied Physics}
{\bf 71} 205 (Heidelberg: Springer)

\item 
Tinkham M 2004
\emph{Introduction to Superconductivity}
(New York: Dover 2nd edn)

\item  
Tisza L 1938
Transport phenomena in helium II
\emph{Nature} {\bf 141} 913

\item  
Tisza L 1947
Theory of liquid helium
\emph{Phys.\  Rev.}\ {\bf 72} 838

\item 
Vitiello S A Runge K J and Kalos M H 1988
\emph{Phys.\ Rev.\ Lett.}\ {\bf 60} 1970

\item 
Vitiello S A Runge K J Chester G V and Kalos M H 1990,
\emph{Phys.\ Rev.}\ B {\bf 42} 228

\item  
Walmsley R H and Lane C T 1958
Angular momentum of liquid helium
\emph{Phys.\ Rev.}\ {\bf 112} 1041.
These authors measure the total change in angular momentum of He~II
from an initial steady state to stationary bucket with zero torque,
and generally find it to be less
than if the entire liquid were initially rotating.
To be consistent with the measurements of Osborne (1950),
there must remain undetected angular momentum.
This, the authors surmise, must be
in the form of `macroscopic rotational states of long lifetime'
as in `the Onsager-Feynman vortex model for He II'.
Alternatively, since the superfluid decouples
from the finally stationary bucket and normal fluid,
it could continue in rigid  rotation
(or in more complicated patterns) without detection,
with a lifetime dependent on the reaction rate.


\item  
Whitlock P A and Panoff R M 1987
\emph{Can.\ J.\ Phys.}\ {\bf 65} 1409

\item  
Wikipedia 2025
Hagen-Poiseuille equation.
Navier-Stokes equations.  (Accessed 1 May, 2025)

\end{list}



\appendix
%
\section{A Little More Thermodynamics}
\label{Sec:ds/de-etc}
\renewcommand{\theequation}{\Alph{section}.\arabic{equation}}
%

 The entropy density can be written
as a function of the bare energy density and the number densities
at that position and time,
$\sigma({\bf r},t) = \sigma^{(0)}(\varepsilon^{(0)},n_0,n_*)$.
The derivatives are
\begin{equation}
\left( \frac{\partial \sigma}{\partial \varepsilon}
\right)_{\underline n,\underline {\bf p},\psi}
=
\frac{\partial \sigma^{(0)}}{\partial \varepsilon^{(0)}}
= \frac{1}{T} ,
\end{equation}

\begin{equation}
\left( \frac{\partial \sigma}{\partial \psi}
\right)_{\varepsilon,\underline n,\underline {\bf p}}
=
\frac{\partial \sigma^{(0)}}{\partial \varepsilon^{(0)}}
\left( \frac{\partial \varepsilon^{(0)}}{\partial \psi}
\right)_{\varepsilon,\underline n,\underline {\bf p}}
= \frac{-n}{T} ,
\end{equation}

\begin{equation}
\left( \frac{\partial \sigma}{\partial {\bf p}_0}
\right)_{\varepsilon,\underline n,{\bf p}_*,\psi}
=
\frac{\partial \sigma^{(0)}}{\partial \varepsilon^{(0)}}
\left( \frac{\partial \varepsilon^{(0)}}{\partial {\bf p}_0}
\right)_{\varepsilon,\underline n,{\bf p}_*,\psi}
= \frac{-{\bf v}_0}{T} ,
\end{equation}
\begin{equation}
\left( \frac{\partial \sigma}{\partial {\bf p}_*}
\right)_{\varepsilon,\underline n,{\bf p}_0,\psi}
=
\frac{\partial \sigma^{(0)}}{\partial \varepsilon^{(0)}}
\left( \frac{\partial \varepsilon^{(0)}}{\partial {\bf p}_*}
\right)_{\varepsilon,\underline n,{\bf p}_0,\psi}
= \frac{-{\bf v}_*}{T} ,
\end{equation}

\begin{eqnarray}
\left( \frac{\partial \sigma}{\partial n_0}
\right)_{\varepsilon, n_*,\underline{\bf p},\psi}
& = &
\frac{\partial \sigma^{(0)}}{\partial n_0}
+
\frac{\partial \sigma^{(0)}}{\partial \varepsilon^{(0)}}
\left( \frac{\partial \varepsilon^{(0)}}{\partial n_0}
\right)_{\varepsilon, n_*,\underline{\bf p},\psi}
 \\ & = &\nonumber
\frac{-\mu^{(0)}_0-\psi+mv_0^2/2}{T}
=
\frac{-\mu + mv_0^2}{T},
\end{eqnarray}
and
\begin{eqnarray}
\left( \frac{\partial \sigma}{\partial n_*}
\right)_{\varepsilon, n_0,\underline{\bf p},\psi}
& = &
\frac{\partial \sigma^{(0)}}{\partial n_*}
+
\frac{\partial \sigma^{(0)}}{\partial \varepsilon^{(0)}}
\left( \frac{\partial \varepsilon^{(0)}}{\partial n_*}
\right)_{\varepsilon, n_0,\underline{\bf p},\psi}
 \\ & = &\nonumber
\frac{-\mu^{(0)}_*-\psi+mv_*^2/2}{T}
=
\frac{-\mu + mv_*^2}{T}.
\end{eqnarray}

\end{document}